\PassOptionsToPackage{unicode}{hyperref}
\PassOptionsToPackage{hyphens}{url}
\PassOptionsToPackage{dvipsnames,svgnames,x11names}{xcolor}
\documentclass[
  10pt,
  twocolumn]{article}
\usepackage{amsmath,amssymb}
\usepackage{iftex}
\ifPDFTeX
  \usepackage[T1]{fontenc}
  \usepackage[utf8]{inputenc}
  \usepackage{textcomp} % provide euro and other symbols
\else % if luatex or xetex
  \usepackage{unicode-math} % this also loads fontspec
  \defaultfontfeatures{Scale=MatchLowercase}
  \defaultfontfeatures[\rmfamily]{Ligatures=TeX,Scale=1}
\fi
\usepackage{lmodern}
\ifPDFTeX\else
\fi
\IfFileExists{upquote.sty}{\usepackage{upquote}}{}
\IfFileExists{microtype.sty}{% use microtype if available
  \usepackage[]{microtype}
  \UseMicrotypeSet[protrusion]{basicmath} % disable protrusion for tt fonts
}{}
\makeatletter
\@ifundefined{KOMAClassName}{% if non-KOMA class
  \IfFileExists{parskip.sty}{%
    \usepackage{parskip}
  }{% else
    \setlength{\parindent}{0pt}
    \setlength{\parskip}{6pt plus 2pt minus 1pt}}
}{% if KOMA class
  \KOMAoptions{parskip=half}}
\makeatother
\usepackage{xcolor}
\usepackage{graphicx}
\makeatletter
\newsavebox\pandoc@box
\newcommand*\pandocbounded[1]{% scales image to fit in text height/width
  \sbox\pandoc@box{#1}%
  \Gscale@div\@tempa{\textheight}{\dimexpr\ht\pandoc@box+\dp\pandoc@box\relax}%
  \Gscale@div\@tempb{\linewidth}{\wd\pandoc@box}%
  \ifdim\@tempb\p@<\@tempa\p@\let\@tempa\@tempb\fi% select the smaller of both
  \ifdim\@tempa\p@<\p@\scalebox{\@tempa}{\usebox\pandoc@box}%
  \else\usebox{\pandoc@box}%
  \fi%
}
\def\fps@figure{htbp}
\makeatother
\NewDocumentCommand\citeproctext{}{}

\makeatletter
 \let\@cite@ofmt\@firstofone
 \def\@biblabel#1{}
 \def\@cite#1#2{{#1\if@tempswa , #2\fi}}
\makeatother
\newlength{\cslhangindent}
\newlength{\csllabelwidth}
\newenvironment{CSLReferences}[2] % #1 hanging-indent, #2 entry-spacing
 {\begin{list}{}{%
  \setlength{\itemindent}{0pt}
  \setlength{\leftmargin}{0pt}
  \setlength{\parsep}{0pt}
  \ifodd #1
   \setlength{\leftmargin}{\cslhangindent}
   \setlength{\itemindent}{-1\cslhangindent}
  \fi
  \setlength{\itemsep}{#2\baselineskip}}}
 {\end{list}}
\usepackage{calc}

\newcommand{\CSLLeftMargin}[1]{\parbox[t]{\csllabelwidth}{\strut#1\strut}}
\newcommand{\CSLRightInline}[1]{\parbox[t]{\linewidth - \csllabelwidth}{\strut#1\strut}}

\usepackage{dcolumn, rotating, longtable, float, array, tabularx, inputenc}

\graphicspath{{figures/}}
\usepackage[margin=2.5cm]{geometry}
\usepackage{setspace}
\usepackage{bookmark}
\IfFileExists{xurl.sty}{\usepackage{xurl}}{} % add URL line breaks if available
\hypersetup{
  pdftitle={Identifying companies and financial actors exposed to marine tipping points},
  colorlinks=true,
  linkcolor={blue},
  filecolor={Maroon},
  citecolor={blue},
  urlcolor={blue},
  pdfcreator={LaTeX via pandoc}}

\title{Identifying companies and financial actors exposed to marine
tipping points}
\author{\small Juan C. Rocha\textsuperscript{1,2}, Jean-Baptiste
Jouffray\textsuperscript{1,3}, Frida Bengtsson\textsuperscript{1},
Bianca-Ioana Voicu\textsuperscript{1}, Paula A.
Sánchez\textsuperscript{1,4}, Victor Galaz\textsuperscript{1}\\
\footnotesize \textsuperscript{1}Stockholm Resilience Centre, Stockholm
University, 10691, Stocholm, Sweden\\
\footnotesize \textsuperscript{2}The Anthropocene Laboratory, The Royal
Swedish Academy of Sciences, Stockholm, Sweden\\
\footnotesize \textsuperscript{3}Stanford Center for Ocean Solutions,
Stanford, CA 94305, USA\\
\footnotesize \textsuperscript{4}Leibniz-Centre for Agricultural
Landscape Research (ZALF)\\
\small \texttt{\href{mailto:juan.rocha@su.se}{\nolinkurl{juan.rocha@su.se}}}}
\date{}

\begin{document}
\maketitle

\begin{abstract}
  \textit{Climate change and other anthropogenic pressures are likely to induce tipping points in marine ecosystems, potentially leading to declines in primary productivity and fisheries. Despite increasing attention to nature-related financial risks and opportunities within the ocean economy, the extent to which these tipping points could affect investors has remained largely unexplored. Here we used satellite data to track fishing vessels operating in areas prone to marine regime shifts, as identified by their loss of resilience and vulnerability to marine heatwaves, and uncovered their corporate beneficial owners and shareholders. Despite some data gaps, we identified key countries, companies, and shareholders exposed to tipping risk. We also outline the potential challenges and opportunities that these actors may face if marine ecosystems shift to less productive states.}
\end{abstract}

\section{Introduction}\label{introduction}

Marine ecosystems are expected to undergo large biological
reorganizations that can affect primary productivity and other ecosystem
services\textsuperscript{1}. These abrupt and nonlinear changes, known
as regime shifts, pose major challenges for managers as they are
difficult to predict and reverse\textsuperscript{2}. For example, the
collapse of Newfoundland cod in the 1990s is estimated to have caused
large economic losses, over 40,000 job losses, and the crash of an
iconic industry and cultural practice in the region\textsuperscript{3}.
Likewise, kelp forests have been observed to transition to urchin
barrens, resulting in the loss of nursery habitats for commercially
important fish species\textsuperscript{4,5}. Similarly, over 500 cases
of oxygen-depleted zones have been reported around the globe, leading to
harmful algae blooms and mass mortality events in fish which impact
human health, local livelihoods and food provision\textsuperscript{6,7}.

Exposure to regime shifts is mediated by pressures on marine ecosystems,
such as climate change, overfishing, or pollution\textsuperscript{1}.
Marine heatwaves are expected to increase in frequency and
intensity\textsuperscript{8,9}, with stark economic consequences for
companies, regions and nations that depend on ocean
productivity\textsuperscript{10}. For example, the Gulf of Alaska
experienced a heat anomaly known as the Blob from 2014 to 2019, causing
an estimated US\$24M loss in a fishery valued at US\$103M annually.
Similarly, harmful algae blooms led to US\$40M loss in tourist spending
along the coast of Washington in 2015\textsuperscript{10}. Other areas
highly vulnerable to heatwaves include the tropical Pacific Ocean, the
Caribbean Sea, northern Australia, the Eastern China Sea and the Western
Pacific Ocean\textsuperscript{10}.

Forecasting the ecological and economic consequences of marine regime
shifts is a difficult endeavor due to uncertainty. Increasing sea
surface temperature can destabilize upwellings -- currents that bring
nutrient-rich water to the ocean's surface -- and affect some of the
world's most productive fisheries\textsuperscript{11}. While
insufficient nutrients can reduce primary productivity and cascade up
through marine food webs, an excess can cause coastal eutrophication and
hypoxia\textsuperscript{6}, with detrimental consequences to fish
stocks. Climate change is expected to reduce fish biomass by 3 to 22\%
in tropical and temperate areas by the end of the
century\textsuperscript{12}. Overfishing and nutrient pollution from
agriculture and urban waste can exacerbate this impact. A combination of
oceanographic and climatic factors heightens the risk of regime shifts,
putting both ecosystems and human livelihoods at
risk\textsuperscript{2,13}. Yet, observing and measuring tipping points
is a challenging task typically masked by low signal to noise
ratios\textsuperscript{14}. However, proactive management strategies
that account for these risks could help mitigate exposure to tipping
points\textsuperscript{2}.

Companies and financial institutions are key actors that contribute to
the extraction of marine resources, while also being affected
economically by changes in marine ecosystems. About 35.4\% of fish
stocks globally are currently depleted or
overexploited\textsuperscript{15}, with only 13 companies controlling
around 16\% of wild seafood capture\textsuperscript{16,17}. High market
concentration, a form of inequality, means that a few actors can have
disproportional effects on the system for good or
bad\textsuperscript{18}. Fishing companies contribute to ecological
degradation through harmful practices like
overfishing\textsuperscript{19}, destructive fishing
methods\textsuperscript{20}, and non-compliance with
regulations\textsuperscript{21,22}. By removing biomass from the ocean,
fishing has historically prevented the sequestration of 21.8M tons of
carbon\textsuperscript{15,23}. In addition, the industry's return on
investment is becoming increasingly risky. Modern fishing fleets travel
twice as far but catch only one-third of what they used to per kilometre
travelled\textsuperscript{24}, it is estimated that 54\% of high seas
fishing would be unprofitable without public
subsidies\textsuperscript{15,25}. Companies doing extractive activities
other than fisheries are also projected to have impacts from climate
change from extreme weather events and sea level
rise\textsuperscript{26,27}

\begin{figure*}[ht]
\centering
\includegraphics[width = 7in, height = 7in]{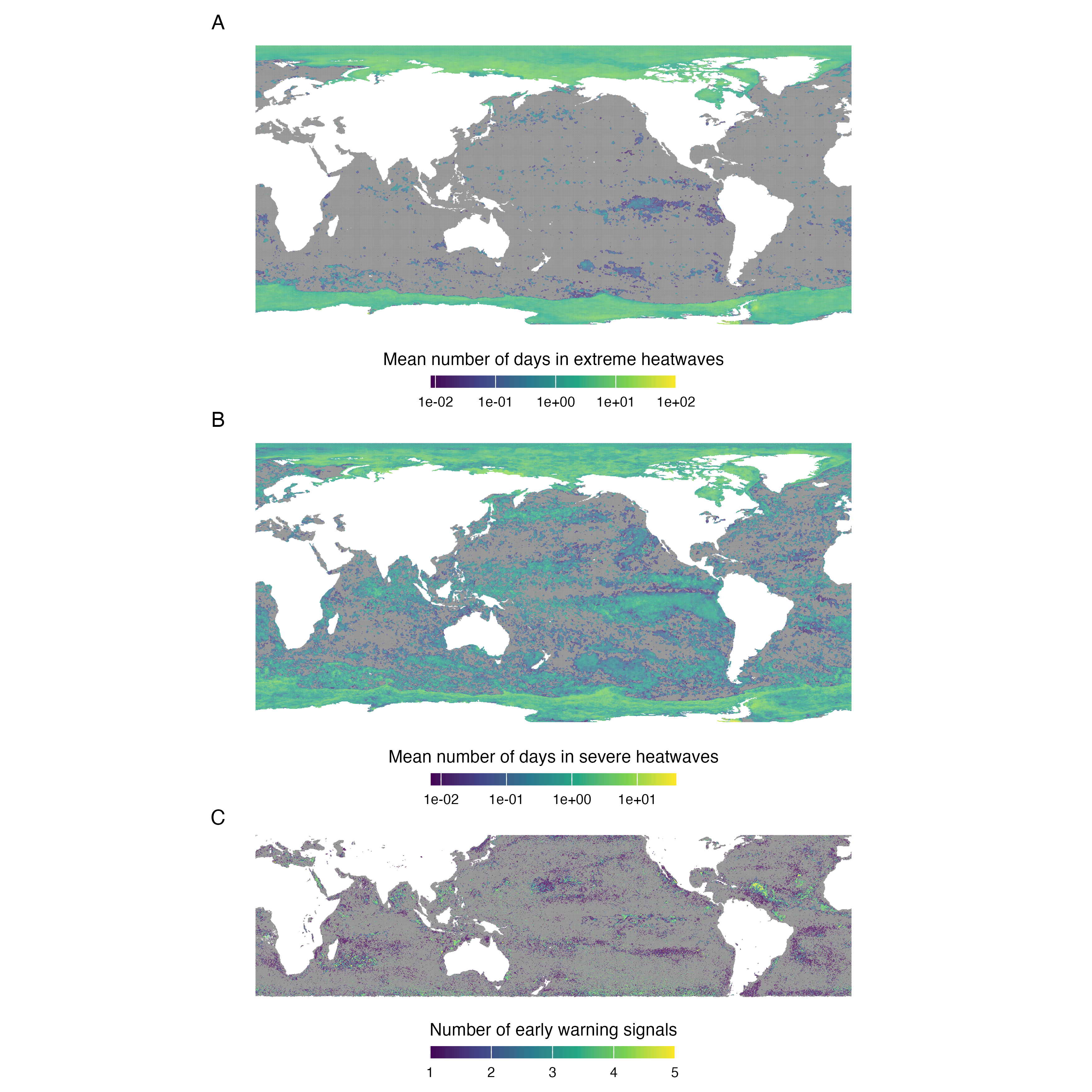}
\caption{\textbf{Areas at risk of tipping points}. We computed extreme and severe heatwave events based on historical records of sea surface temperature (A, B). Areas in colour are places where the probability is non-zero, expressed as the mean number of days spent on a heatwave event. Ocean colour data was used to compute early warning signals of critical transitions (C). Note that there are no warnings computed for higher latitudes due to missing data. For the analysis of areas prone to marine regime shifts, we kept areas where the probability of extreme and severe heatwaves is non-zero, or where at least three early warning signals were found (see Methods). Supplementary figure \ref{fig:rsdb} provides additional empirical evidence of the occurrence of regime shifts by mapping 1448 case studies coded through literature review.} 
\label{fig:maps}
\end{figure*}

It should be in the best interest of companies and their shareholders to
act as stewards of the seas and maintain a productive and healthy
ocean\textsuperscript{28,29}. Financial actors -- such as banks, asset
managers, and pension funds -- could play an important role in
conservation by deciding not to finance economic activities that alter
the ecological integrity of the biosphere\textsuperscript{30--33}. For
instance, a small set of financial actors have an increasing influence
over climate stability through their equity holdings in companies
operating in the Amazon and boreal forests\textsuperscript{30}.
Likewise, large asset managers and other financial institutions invest
in companies operating in areas vulnerable to the emergence of
infectious diseases\textsuperscript{31}. This leaves them exposed to the
impacts of their investments\textsuperscript{34}.

\begin{figure*}[ht]
\centering
\includegraphics[width = 7in, height = 7in]{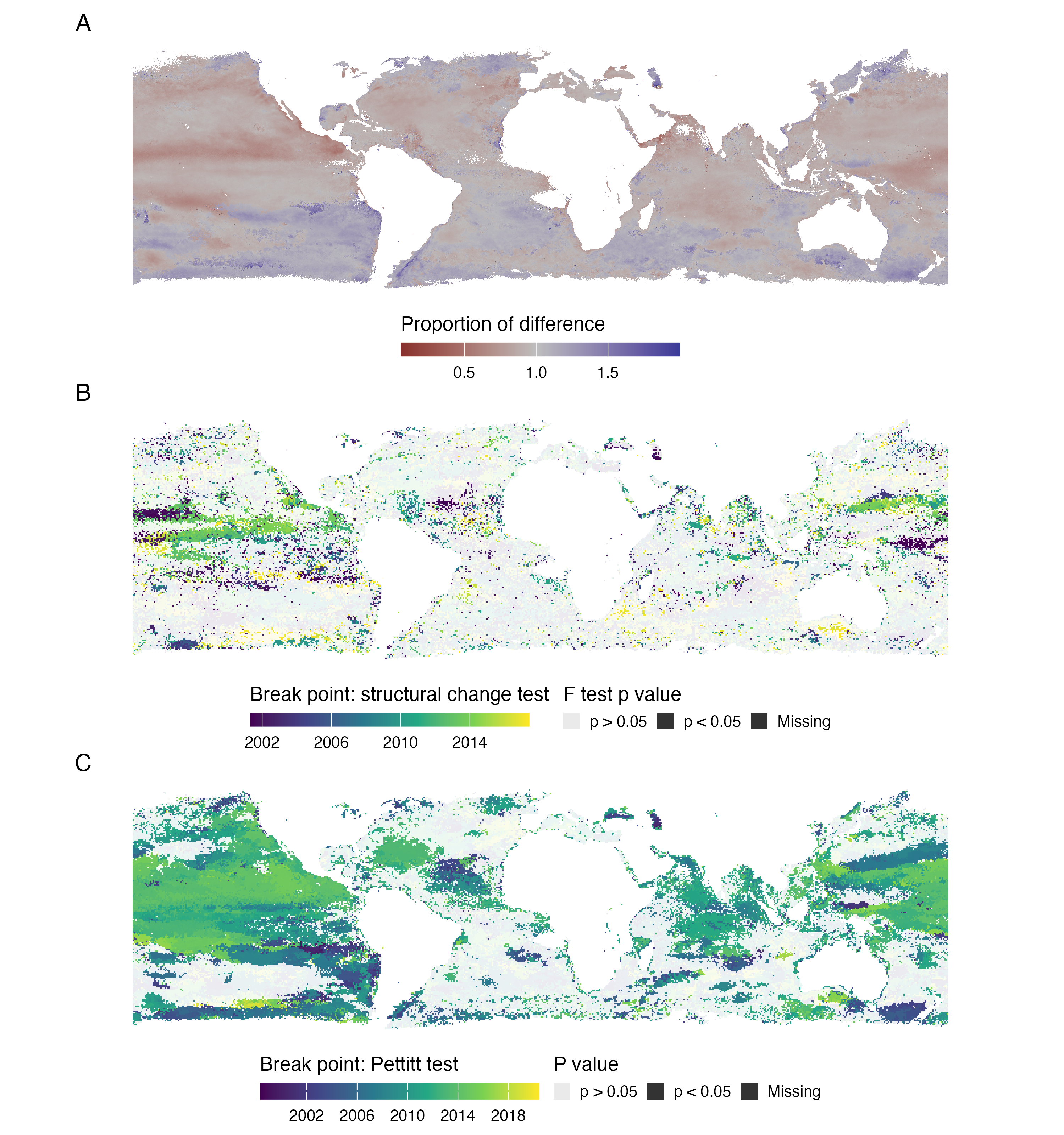}
\caption{\textbf{Abrupt changes in primary productivity}. We used the structural change and Pettitt test to investigate abrupt shifts in Chlorophyll A data and computed the proportion of change before and after the shift (the mean after over the mean before). If the proportion is one there was no change in means, if the proportion is < 1 it means an abrupt decline (e.g. 0.5 is a 50\% decline), or an increase if >1 (A). 16\% of the oceans show significant abrupt changes with the structural change test (B) while 50\% of the oceans show abrupt changes in primary productivity with the Pettitt test (C). Supplementary Figure \ref{fig:slopes} shows a similar test for linear trends in primary productivity.} 
\label{fig:break_point}
\end{figure*}

Climate-change-driven extreme weather events are expected to cost up to
USD 14 trillion annually by 2100\textsuperscript{35}, directly
threatening the solvency of insurers and reinsurers\textsuperscript{36}.
Investors and asset managers face climate exposure being at risk of
stranded assets, changes in commodity prices, scarcity of resources,
disruption of supply chains, and infrastructure
damages\textsuperscript{37}. In response, financial actors are
increasingly expected to disclose their risks through initiatives such
as the Task Forces on Climate- and Nature-Related Financial Disclosures
(TCFD and TNFD, respectively) or the EU non-financial reporting
directive.

\begin{figure*}[ht]
\centering
\includegraphics[width = 6.5in, height = 6in]{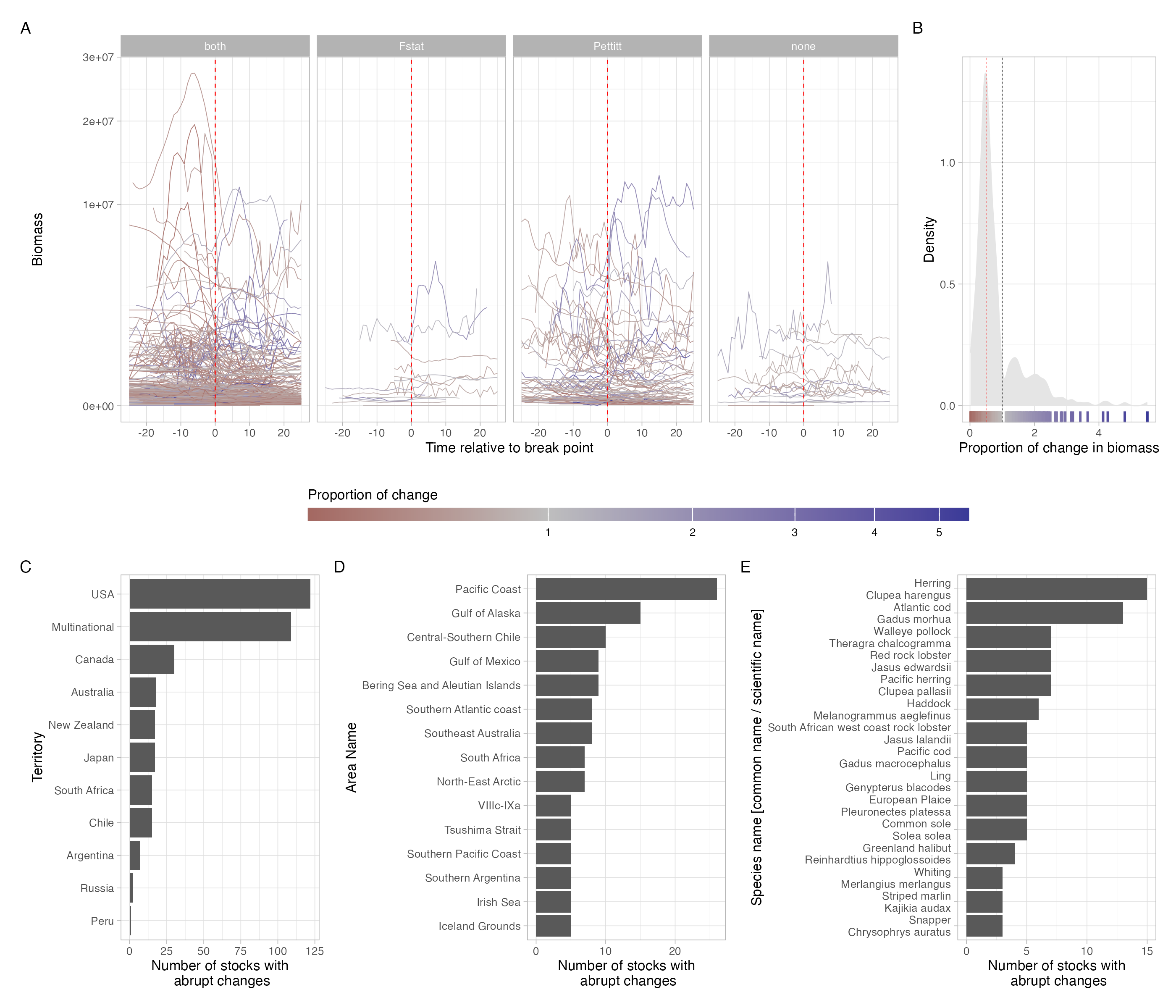}
\caption{\textbf{Abrupt changes in fish biomass}. We tested abrupt changes in fish biomass data from the RAM legacy database using the Pettitt and structural change tests. We found 252 stocks having abrupt changes detected by both tests, 14 only by the structural change F statistic, 83 only by the Pettitt test, and 20 with no detection (A). The mean decline for all significant changes was 50\% of the biomass with respect to the mean before the abrupt change (B). The bottom panel lists the top territories most affected (C), the geographic areas most impacted (D), and the species with most abrupt changes. Supplementary Figure \ref{fig:fish_stocks} complements the analysis by comparing the timing of the shifts between tests and a linear test.} 
\label{fig:stocks}
\end{figure*}

Analyses of financial exposure to biodiversity loss and ecosystem
degradation are mostly focused on terrestrial ecosystems, with little
research addressing ocean-related financial risks\textsuperscript{38}.
Yet, fishing companies, their shareholders, and the countries they
operate in are exposed to potentially irreversible changes and
consequent declines in ecosystem services, with consequences on the
economy, investments and livelihoods. However, being exposed to regime
shifts also means being in a position to take actions that reduce
anthropogenic pressures that could drive the system across a potential
tipping point. In this paper, we asked who are these corporate and
financial actors exposed to marine tipping points?

\begin{figure*}[ht]
\centering
\includegraphics[width = 6.5in, height = 4in]{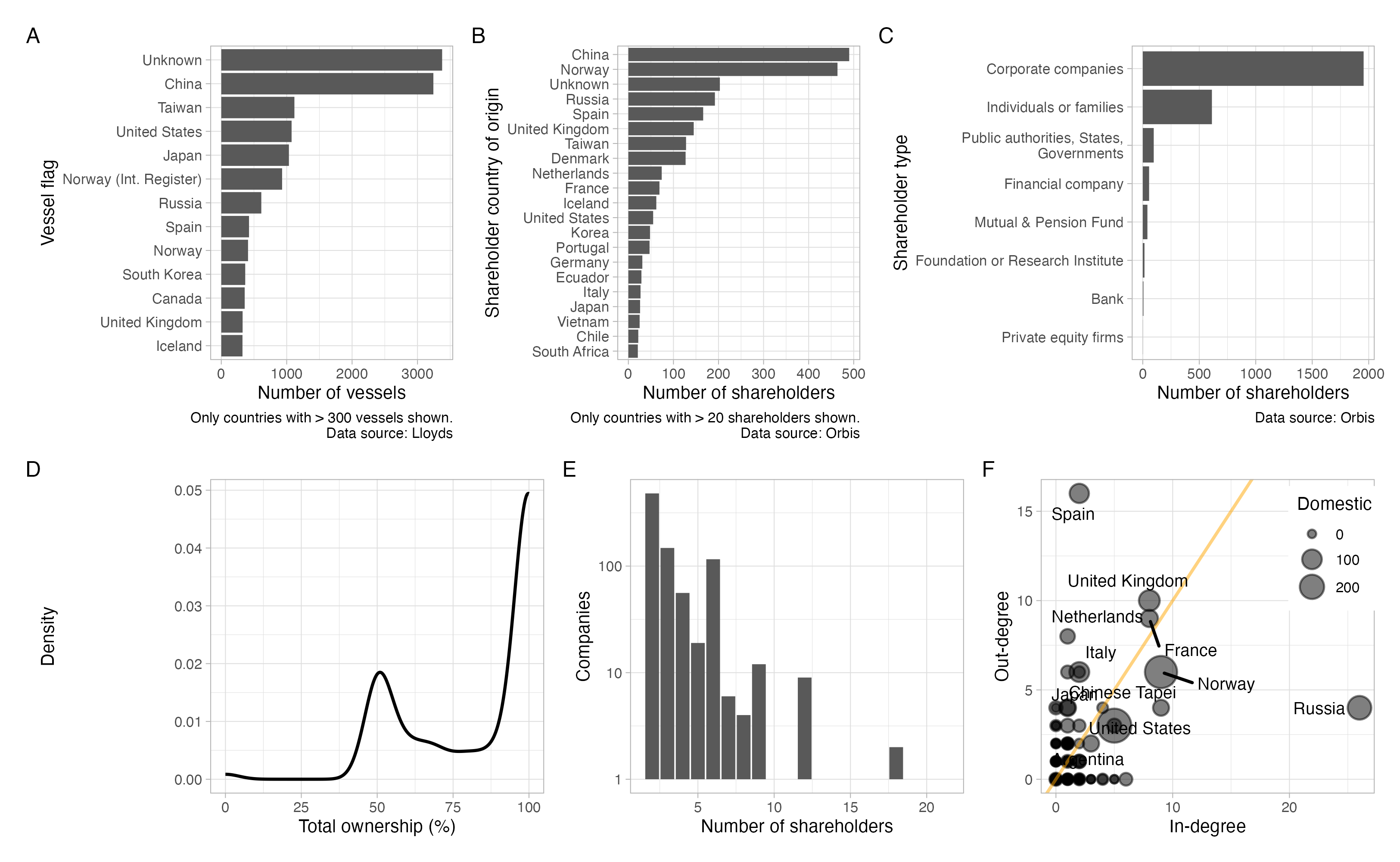}
\caption{\textbf{Descriptive statistics} Number of vessels per flag reported (A), shareholders' country of origin (B), shareholder type (C), and the probability density of their ownership shares (D). Most fishing companies analyzed have only a few shareholders (E). Countries ranked by their number of foreign investments (out-degree) and the number of investments received (in-degree), the size of the dot is proportional to the number of domestic investments (F). Countries in the yellow line have the same number of incoming and outgoing links in the network, above the line are countries with disproportionally more investments, and below countries which receive the investments.}
\label{fig:shrs}
\end{figure*}

\section{Methods}\label{methods}

\textbf{Areas at risk:} We identified areas of the world vulnerable to
marine regime shifts. We approximated these areas as places currently
under the strong influence of marine heatwaves or already showing
symptoms of resilience loss. Frölicher et al.\textsuperscript{8} defined
heatwaves as events when sea surface temperature exceeds the 99th
percentile with respect to daily satellite observations for a 30-year
period. Hobday et al.\textsuperscript{39} further classified heatwave
categories based on the 90th percentile, where moderate heatwaves are
between 1x and 2x the 90th percentile, strong heatwaves between 2x and
3x, severe heatwaves between 3x and 4x, and extreme heatwaves above 4
times the 90th percentile. We computed the probability of severe and
extreme heat waves based on daily sea surface temperature observations
from the US National Oceanic and Atmospheric Administration (NOAA) using
the optimum interpolation data from 1981 to 2022 at 0.25\(^\circ\)
resolution (Fig \ref{fig:maps}A-B). While several impacts of heatwaves
have been documented in ecosystems and the
economy\textsuperscript{10,39}, a recent global assessment showed no
systematic effect on fish biomass\textsuperscript{40}. However,
heatwaves by themselves are seldom the sole cause of fish populations
decline, it is often the compound effect of temperature with other
pressures such as low primary productivity or low oxygen
zones\textsuperscript{13}.

Thus we complemented heat waves detection with proxies of stability loss
in primary productivity, which are in turn driven by oxygen depletion,
salinity, and the mixing of the upper layer of the
ocean\textsuperscript{6,7,41}. We used a geographically explicit mapping
of marine ecosystems showing symptoms of resilience
loss\textsuperscript{42}. Resilience is the ability of a system to
withstand disturbances without losing its functions and
identity\textsuperscript{43}. When systems lose resilience and are close
to tipping points, they leave statistical signatures in time series
known as early warning signals\textsuperscript{44}. The map was produced
by computing temporal autocorrelation, standard deviation, skewness,
kurtosis and fractal dimension on Chlorophyll-A concentration, a proxy
of marine primary productivity, using data from the European Space
Agency (weekly observations at 0.25\(^\circ\) resolution from
1998-2018)\textsuperscript{45}. The data are limited by the coverage of
the ocean colour data (the 50\(^\circ\) N-50\(^\circ\) S region),
excluding the Arctic and Antarctic oceans\textsuperscript{42}. For this
paper we used areas with at least 3 early warnings, since two of them
are suitable for bifurcation tipping (autocorrelation and standard
deviation), two of them are suitable for flickering and stochastic
transitions (skewness and kurtosis), and the fractal dimension is robust
against heteroskedasticity. At least three early warnings give us the
chance to pick up more robust signals of the risk of tipping taking into
account different ways of reaching the transition.

\textbf{Evidence of regime shifts:} We documented past regime shifts by
reviewing scientific literature using the framework of the
\href{www.regimeshifts.org}{regime shifts database}\textsuperscript{46}.
In addition, we performed change point analysis, a classical method to
detect regime shifts in time series\textsuperscript{47,48}. We looked
for break points on the Chlorophyll-A concentration data from the
biological pump and carbon exchange process project\textsuperscript{49}
which offers monthly data at 9km pixel resolution, and is derived from
the European Space Agency's Ocean Colour Climate Change initiative. We
also looked at break points on historical fish stock assessment data
from the RAM legacy stock assessment database version
4.44\textsuperscript{50,51}, similar to previous efforts identifying
regime shifts on fish stocks\textsuperscript{48}. Out of the 479 stocks
for which there are time series reported, we assessed only stocks with
more than 25 years with data (N = 373).~ If there is an abrupt decline
in primary productivity, one can expect that the foodweb depending on it
can be prone to similar changes. However, linear changes in primary
productivity can also generate abrupt decline on fish stocks. For that
reason we also calculated the linear trend on Chlorophyll-A time series
to show areas of the world with significant changes by using the Kendall
tau statistic. For break point detection we used the Pettitt test of
change in means\textsuperscript{52}, and the change point F test
statistic for change on residuals\textsuperscript{53}.

\textbf{Fishing pressure:} We identified fishing vessels operating in
pixels showing a high probability of heatwaves or symptoms of resilience
loss. We used the dataset published by Kroodsma et
al.\textsuperscript{54} who identified \textgreater70,000 industrial
fishing vessels in the size range from 6 to 146m in length, thanks to an
automatic identification system (AIS). Vessels need to broadcast their
position on the AIS to avoid potential collisions, so the data offers
individual vessel trajectories from 2012-2020 (fishing effort dataset
from Global Fish Watch version 2.0). We extracted vessels whose fishing
activities were recorded in areas with high probabilities of heatwaves
(severe or extreme events) or high probability of regime shifts. We only
include pixels with at least three different early warning signals of
resilience loss, that way we include multiple routes of tipping
(e.g.~noise-induced, flickering, not only the traditional slowing down),
and keep pixels for which multiple lines of evidence suggest loss of
resilience.

\textbf{Owners and shareholders:} We identified the individual and
company beneficial owners of the vessels using the maritime mobile
service identity to match the Global Fishing Watch data with vessel
owners in the Lloyds database
(\href{http://www.seasearcher.com/}{www.seasearcher.com}). We matched
13,537 records, but not all of them had available ownership information.
We also used the dataset curated by Carmine et al.\textsuperscript{55}
where vessels fishing on the high seas were manually annotated. We added
companies whose vessels matched our selection of vessels fishing in
areas with marine tipping point risk. To identify shareholders and
global ultimate owners of the companies responsible for fishing
activities, we used Orbis, a database of companies around the world with
coverage of \textgreater{} 489 million companies. Where possible, we
extracted data about their shareholder structure (e.g.~direct and total
ownership, shareholder type), annual revenues, number of employees, and
shareholder's countries of operation. Unfortunately, Orbis and Lloyds
are not open access data sources and their terms of service forbids
sharing the raw accessed data with third parties. Both of them were
accessed through research subscriptions by the home institution.

\textbf{Network models:} We constructed an international network of
foreign investments using the countries responsible for the fishing
activities (vessel flags) and the shareholder's country of origin. This
network representation allowed us to explore the likelihood of
investments (probability of a link) since investing in a particular
country is not independent of network structure (e.g.~investments
already present in the network, or individual features of the countries
involved). We used governance data from the World Bank (control of
corruption, government effectiveness, political stability, regulatory
quality, rule of law, and voice accountability) to investigate if
foreign investments in fishing activities exposed to marine tipping
points tend to occur in countries with lower effectiveness and weaker
rule of law, as previously reported in the context of emerging zoonotic
diseases\textsuperscript{31}. To test this hypothesis, we used an
exponential random graph model\textsuperscript{56} where nodes are
countries and links are investment flows between shareholders and
companies running fishing operations.

\section{Results}\label{results}

\begin{figure*}[ht]
\centering
\includegraphics[width = 6.5in, height =3in]{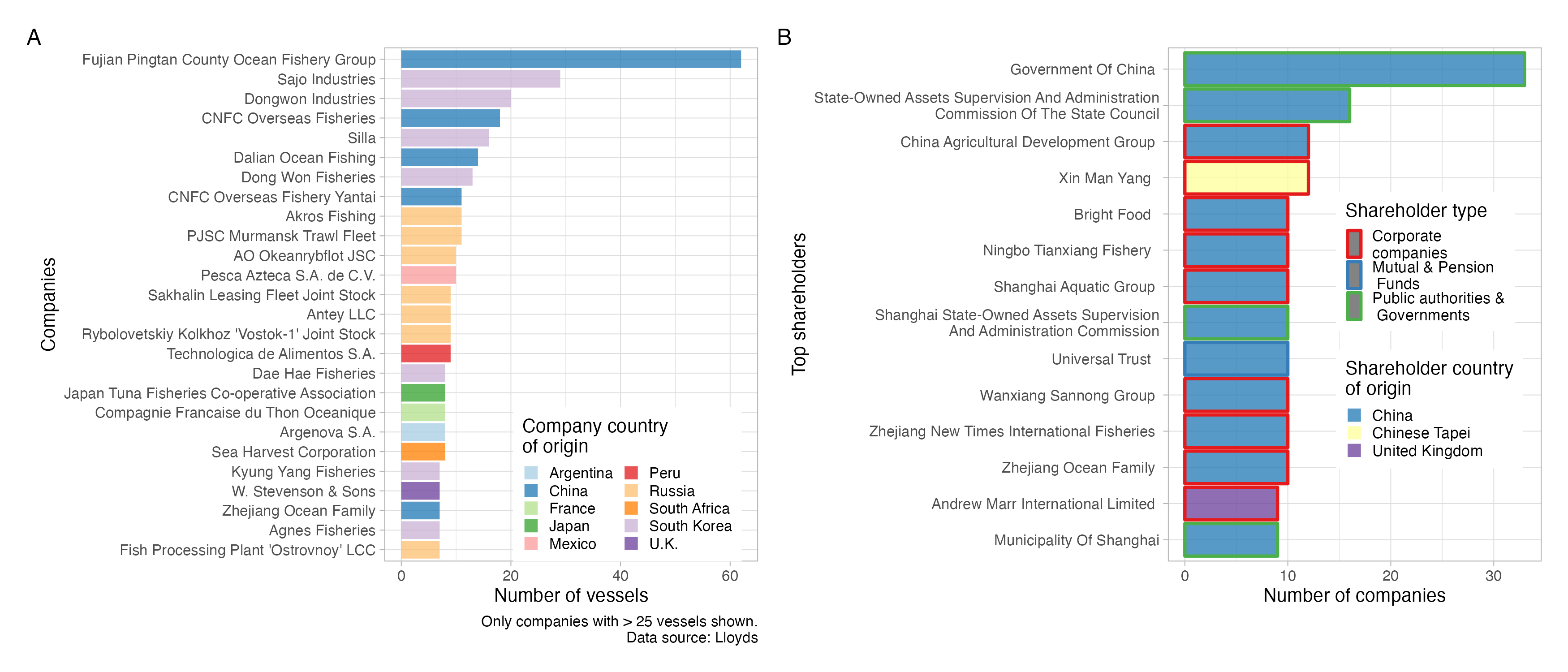}
\caption{\textbf{Actors exposed to marine tipping} Top companies (A) and shareholders (B) exposed to areas with high risk of regime shifts. Colours correspond to country and countours to shareholder types (B). The rank for global ultimate owners is reported in Fig. \ref{sm:guos}}
\label{fig:top}
\end{figure*}
\begin{figure*}[ht]
\centering
\includegraphics[width = 6.5in, height = 3in]{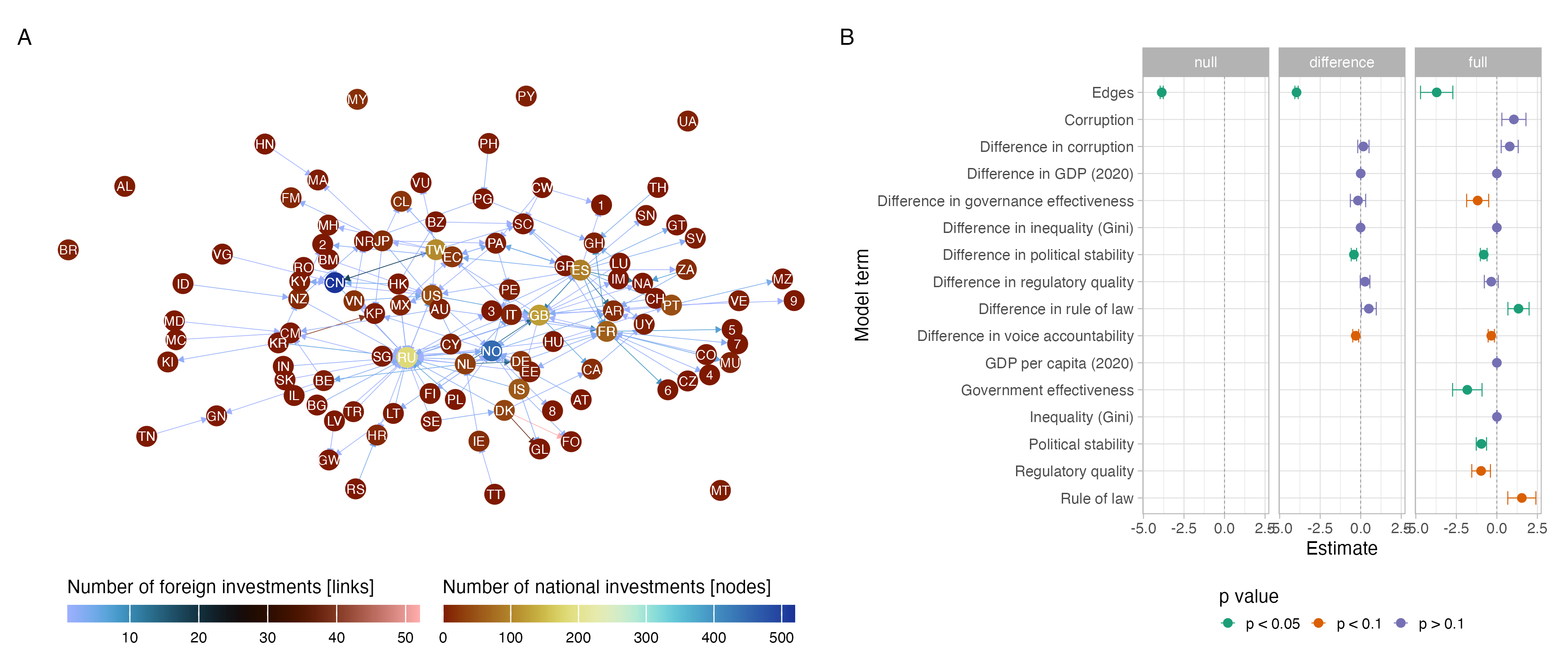}
\caption{\textbf{Network of shareholder investments} While most investments are domestic, a network of foreign investments emerge featuring European countries, China and Russia as central actors (A). Results from exponential random graph models on the World Bank governance idicators (B) reveals that investments tend to occur when differences in governance efficiency and political stability exist. Two letter codes were used to identify countries (ISO 3166-1 alpha 2 standard), for territories without a code we used numbers as follows: Curaçao: 1, Cook Islands: 2, Falkland Islands: 3, French Southern Territories: 4, Mayotte: 5, Reunion: 6, St. Pierre and Miquelon: 7, St. Helena: 8, and Azores: 9.}
\label{fig:net}
\end{figure*}

We found that 30\% of global oceans have presented extreme heat waves,
and 66\% severe events (Fig \ref{fig:maps}). Most of the extreme events
are located in high latitudes in the Arctic and Antarctic oceans. Severe
events are also common in the tropical Pacific ocean (related to ENSO),
coastal California, and the Indian ocean. Upwelling systems in the coast
of California, Benguela or Canary islands show at least three early
warning signals of the proximity of instabilities. 24\% of the area
analysed with early warning signals show symptoms of resilience
loss\textsuperscript{42}. The tropical north Atlantic is also a hotspot
of resilience loss, a place where previous work has documented increase
in summer time stratification of the water column driven by sea surface
temperature and salinity\textsuperscript{41}. With stronger
stratification there is less mixing of water column and nutrient
cycling, leading to a decline in primary productivity. We detect such
signal in the linear decline of primary productivity (Fig
\ref{fig:slopes}).

Our change point analysis shows that there has been abrupt changes in
primary productivity over time (Fig \ref{fig:break_point}). The extend
of the area varies depending on the statistical test used. The Pettitt
test looks for changes in means, and detects significant changes in 50\%
of the oceans with break points in early 2000s in the south Pacific, and
more recent changes in the rest of the world. The structural change test
is more restrictive, seeking changes on the residuals of a lagged
autoregressive model. We find again abrupt changes in the central
Pacific, the tropical Atlantic, the coasts of California, and India;
adding up to 16\% of the oceans. For the test that looks if there is a
decline on Chlorophyll-A regardless if it is abrupt or smooth (the slope
of a linear regression), we found that 77.3\% of the ocean area (50N-50S
region) has a significant kendall tau, and for 41.9\% of the area
primary productivity is declining (Fig \ref{fig:slopes}).

Similarly, for the stock assessment data we found that 353 fish stocks
show significant abrupt changes of which 271 are declining biomass while
82 are increasing (Fig \ref{fig:stocks}). We find agreement between the
two methods used for change point detection (256 stocks), but some
differences in the time predicted. The mean decline between the mean
before and the mean after the break point is of 50\% of the biomass.
Qualitatively, our results align with previous studies using different
methods, smaller sample sizes but stricter thresholds for
collapse\textsuperscript{48,57--59}, where the proportion of stocks
collapsed are between 25\% and 50\%. Here we find that 119 stocks
collapsed below 50\% their historical average (31\%), while 29 stocks
(7\%) collapsed below 15\% their historical record (Fig
\ref{fig:stocks}B). Regardless if the changes are abrupt, the Sen slope
test supports the direction of change identified through the change
point analysis (Fig \ref{fig:slopes}). The countries with more fish
stocks showing abrupt changes are USA, Canada and Australia, although
many occur in international waters; the fishing grounds where most
abrupt changes reported are the Pacific coast, the Gulf of Alaska,
central and southern Chile and the Gulf of Mexico. The species most
impacted are herring, Atlantic cod, walley pollock, red rock lobster and
pacific herring (Fig \ref{fig:stocks}).

Out of the 114,191 vessels reported in the Global Fish Watch database,
we found that 15 \% of all vessels tracked (N = 16,878) fish in areas
prone to marine tipping points (Fig \ref{fig:maps}). The bulk of these
vessels do not report a flag or country of registry; however, for those
who do, the largest fleets are associated to China, the United States,
Japan and Norway (Fig \ref{fig:shrs}A, see Fig \ref{sm:vessels} for
vessels types ). China's fleet is at least three times larger than the
next larger fleet, making it a dominant actor. Out of the 16,878
vessels, we identified 2,981 companies, for 855 of them we identified
1,825 unique shareholders and 2,783 ownership relationships between
shareholders and companies. Most of the shareholders are based in
Norway, China, Russia, Spain, the United Kingdom, and Denmark (Fig
\ref{fig:shrs}B); the most common shareholder types are corporate
entities and individual families.

The network of ownership and shareholder investments is relatively
sparse, lacking the presence of major financial giants common to
multiple companies. Unlike previous studies that identified ``keystone
actors'' -- central corporations and financial entities with
disproportionate influence in the market\textsuperscript{16} and
potential exposure to tipping elements of the Earth
system\textsuperscript{30}-- our findings reveal a more diffuse network
with no central actors. This is similar to the pattern found by Jouffray
et al.~when looking at shareholders of publicly listed seafood
companies\textsuperscript{33}. One possible explanation is the ownership
structure where most fishing companies in our dataset are owned by only
a few shareholders who in turn own on average half, if not all, of the
company's shares. (Fig. \ref{fig:shrs}D, E). This is reflected by a
large proportion of our shareholders being individuals, families, or
small-scale businesses (Fig \ref{fig:shrs}C). Another possibility is
that keystone actors would emerge in the network only when considering
not just vessel owners, but also retailers and traders involved in the
supply chain.

Despite the sparsity of the ownership network, there were actors with
disproportional weights on the network (Fig \ref{fig:top}). The top
company in our data set controlled over 60 vessels, and most top
companies operated in China, South Korea or Russia. The top shareholders
were also predominantly based in China with several of them being
corporate companies (e.g.~all companies that are not banks, financial or
insurance companies). However, in terms of the number of companies
invested in, public authorities and government offices emerged as the
most important actors. This emphasizes the key role that public
authorities play in not only regulating fishing operations, but also in
investing and subsidizing activities that can be detrimental to the
long-term interest of the citizens they represent. Notably, our findings
show that this influence is not restricted to national governments, and
that regional and city-level authorities could also be key actors in
fisheries management.

The network of international investments revealed a different set of
relevant actors (Fig \ref{fig:net}). Most shareholders were based in
Norway but their investments were largely domestic and focused on
Norwegian fishing companies. International investments were dominated by
Spain, the UK and France as countries where investments originated,
while Russia seemed to disproportionately receive investments (Fig
\ref{fig:net}, Fig \ref{fig:shrs}F). Using this network of international
investment we asked what features increased the likelihood of an
investment.

To answer this question we fit three exponential random graph models of
the network as response variable and governance indicators as
explanatory variables (Fig \ref{fig:net}B). The null model simply
assumes that investment relationships form randomly. The difference
model only takes into account the difference in governance statistics
between the country where the investor is based minus the statistic
where the fishing company is located. The full model uses both the
differences and the raw statistics per country to predict the likelihood
of investments. The full model outperformed the other two by maximizing
the log likelihood and minimizing the Akaike Information Criterion (see
SM table \ref{tab1}).

We found that government effectiveness was the best predictor of foreign
investments (Fig \ref{fig:net}). Countries with effective governments
tended to invest in governments that have less effectiveness (hence the
positive difference in the full model). We also found a smaller, yet
significant and negative effect on political stability. Countries with
low political stability in our data set invested in companies located in
countries with slightly higher political stability (hence the negative
difference on the full model). The implications of this result is that
the relevant decisions to manage tipping point risk are not only in the
hands of fishing vessels, the companies that own them or their
shareholders. The countries where these activities take place (fishing
and investing) can design regulations that restrain or disincentivise
investments in other regions of the world where fishing activities are
happening in high risk areas.

\begin{figure*}[ht]
\centering
\includegraphics[width = 6.5in, height = 2.5in]{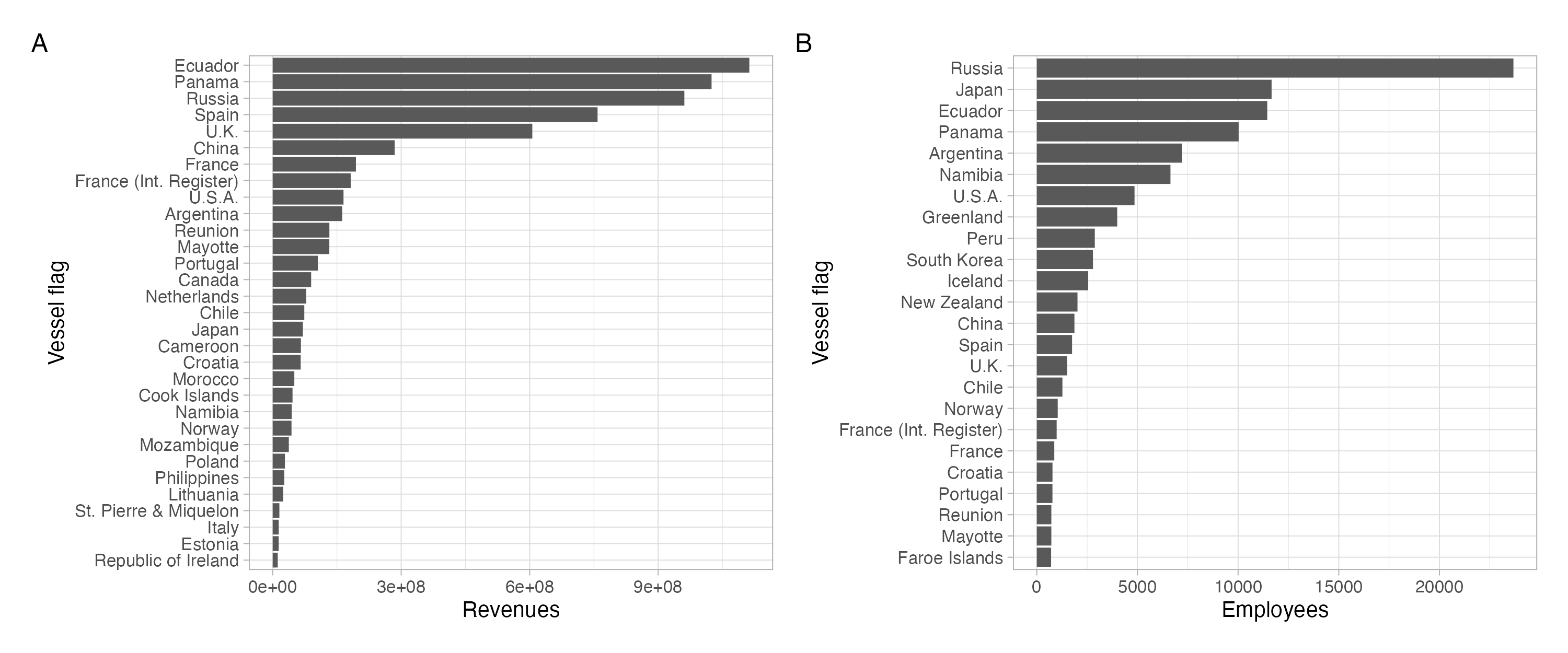}
\caption{\textbf{Potential impacts of marine tipping points}. Countries ranked by the revenues reported in our dataset (A). Only countries with revenues higher than 1M US\$ shown. Countries ranked by workforce employeed (B), only countries with more than 500 employees shown. The data however has high percentage of missing values (\~70\%), so it should be interpreted as an under estimation.}
\label{fig:impacts}
\end{figure*}

Despite data incompleteness, our results show that if marine tipping
points were to synchronize globally, they could impact an industry
generating over 6.6 billion dollars in revenue and affect more than
100,000 jobs (Fig \ref{fig:impacts}). The vessel flags producing the
highest revenues in our data were Ecuador, Panama, Russia, Spain and the
UK, while those employing the largest workforces were led by Russia,
Japan, Ecuador, Panama and Argentina (Fig \ref{fig:impacts}). Vessel
flags are not necessarily where profits go, but data gaps prevented a
more accurate approximation. We were able to retrieve revenue and
employment data for only 2,216 companies, with roughly 70\% of the data
missing values. As a result, our figures likely underestimate the true
potential economic and social impacts of possible declines in primary
productivity from marine heatwaves and marine regime shifts. Much higher
levels of transparency in company reporting are needed to improve these
initial estimates.

\section{Discussion}\label{discussion}

Recent studies showed that heatwaves alone are unlikely to decrease fish
biomass\textsuperscript{40} despite some events have been already
observed\textsuperscript{8,10,39}. Yet, it is often the effect of
compound events or the interaction of multiple drivers that can drive
abrupt fish declines or other marine regime
shifts\textsuperscript{2,13}. Their frequency and intensity are
predicted to increase in the near future with climate change
scenarios\textsuperscript{1,8,9,13}. In this paper, we examined which
corporate and financial actors are exposed to marine tipping points by
identifying vessels and companies operating in areas that are
simultaneously exposed to marine heat waves and instabilities in primary
productivity. We focused on heat waves and changes in primary
productivity because of data availability, acknowledging that marine
ecosystems can tip due to other causes as well\textsuperscript{2}.

Here we reported lists of top countries, companies, and their
shareholders associated with vessels fishing in areas at high risk of
marine regime shifts. In summary, we found that fishing fleets
associated with China, the US, Japan and Norway are particularly
exposed. The shareholders of companies exposed to tipping point risks
are based in Norway, China, Spain, Russia, the UK and Denmark. Spain, UK
and France are home to particularly vulnerable shareholders investing in
companies who fish in high-risk areas, while Russia is headquarters for
fishing companies getting such investments. The implications are that
the management is not only in the hands and decisions of companies.
Countries could also design laws and regulations that restrain or
disincentivise investments in other regions of the world where fishing
activities are occurring in risk areas. We developed a method to track
vessels and companies at risk which depends on a mix of publicly
available\textsuperscript{55} and non-public datasets.

Our method and findings are limited by data completeness and
transparency. For example, corporate ownership often involves shell
companies, offshore entities and indirect holdings that may not be fully
documented in Orbis, or risk misidentifying or omitting ultimate owners.
Another example is vessel flags, many vessels operate with flags of
convenience obscuring the true ownership structure in the data. The
absence of financial data at vessel level further weakens the link
between company ownership and fishing activities.

While transparency and accountability are values sought after by the
financial sector, our analysis highlights the complexity and opacity of
global fisheries, even with access to state-of-the-art
data\textsuperscript{54,55} and databases (e.g.~Lloyds, Orbis). We first
identified close to 17,000 vessels fishing in high-risk areas, but as we
traced ownership up the chain --from vessel owner to company
beneficiaries and shareholder-- data availability dropped by an order of
magnitude at each step. A high proportion of vessels have unidentified
flags (20\%) and lack registered owners (81.41\%). Thus our results are
biased towards the vessels and countries that report ownership data,
that is primarily publicly listed companies operating in legal
fisheries. Further research is needed to correct this
bias\textsuperscript{60}. Nonetheless, our study serves as a benchmark
for what is achievable with current databases and is the first study to
connect ocean tipping points to financial risks.

Another limitation stems from known problems with the AIS data on which
our vessel identification is based. The data is known to be less
reliable before 2017 and not all countries follow the same standards in
reporting. For example, while AIS is primarily used by vessels over 300
gross tons operating beyond national jurisdiction, a country like China
mandates AIS for a broad range of vessels also within the Exclusive
Economic Zone (EEZ). This has led previous studies to restrict their
analyses to the high seas\textsuperscript{55}. Here, we nevertheless
included EEZs in our analysis because they are often prone to marine
heatwaves and regime shifts (Fig \ref{fig:maps}). In fact, over 500
observed hypoxic and anoxic zones --a regime shift that impacts primary
productivity and hence fishing-- have been observed in national
jurisdictions\textsuperscript{6,7}.

Despite these limitations, our study offers key insights for managing
the financial risks associated with marine tipping points. In contrast
to previous work that identified keystone actors dominating the
exploitation of natural resources\textsuperscript{17,30,31}, we found
that the fisheries sector does not seem to be highly concentrated, at
least not at the level of decision-making that is exposed to risk of
regime shifts -- where to go fishing and when. Market concentration
probably occurs further down the supply chain, at the level of
intermediaries\textsuperscript{33}, but this level of aggregation makes
it harder for financial institutions and regulators to discriminate
whether the fish traded comes from areas prone to tipping points.

Financial actors and corporations are increasingly required to report
their climate and nature-related risks, as seen in initiatives like the
Task Force on Climate-Related Disclosures (TCFD), or the European
Corporate Reporting Directive (CSRD). Central to these reporting efforts
is the concept of double materiality: companies need to assess what is
their vulnerability to ecological or climate change; but also how their
activities impact climate change or ecological
degradation\textsuperscript{61}. This feedback loop heightens the risks
faced by both the company and its investors. Our maps of tipping point
risk combined with readily available georeferenced information on
fishing fleet location enable companies and financial investors to
assess their exposure, adjust their fishing areas (companies), and
manage their investment portfolio (investors) to reduce exposure.
Heatwaves and cold spells forecasting tools\textsuperscript{9}, as well
as near-real-time monitoring of ecosystem resilience\textsuperscript{42}
can support adaptive strategies that minimize risk exposure.

When fisheries collapse (e.g.~cod in the 1990s) or when stocks decline
abruptly (e.g.~the Blob in 2014-19), governments often need to step up
and pay the economic and social costs. At the same time, governments are
subsidizing high seas fisheries, without such subsidies 54\% of their
catch would have been unprofitable otherwise\textsuperscript{15}. Our
study shows that governments also have a role to play as shareholders in
some of the biggest fishing fleets. This is particularly evident in
China where national, regional and municipal government organizations
sit on the boards of fishing companies exposed to tipping point risks.
These governmental and public organizations have decision power to
mitigate risk by directing fisheries to less risky grounds or by holding
the fleet accountable for avoiding high-risk areas.

As heatwaves and marine regime shifts unfold, there is an additional
risk that governance frameworks may become outdated. While benthic
species tend to have low mobility and be impacted by these events,
pelagic species can more readily relocate if the area affected is within
their mobility range. The problem arises when a stock quota that was
assigned in an economic zone to a particular set of actors moves to
another economic jurisdiction, generating conflicts between fisheries
and their managing institutions\textsuperscript{62,63}. Many fish stocks
are already expected to move polewards due to climate
change\textsuperscript{64}, and marine regime shifts may exacerbate
these tensions. Governments and fishing authorities should put
mechanisms in place to prevent, mediate or resolve these emerging
conflicts, as non-linear environmental changes can not only alter the
``rules of the game'' but also shift the playing field.

\section{Conclusion}\label{conclusion}

In this paper, we asked who are these corporate and financial actors
exposed to marine tipping points? We first identified areas of the ocean
where tipping points can manifest because they are already showing a
high intensity of extreme and severe heat wave events, or they are
already showing statistical signatures of resilience loss in primary
productivity. We found that about 30\% of the oceans are at risk of
regime shifts (extreme marine heat waves and signals of instability in
primary productivity), and provided evidence of past regime shifts from
literature reviews, and abrupt change observed in primary productivity
and stock assessments data. 16\% of the oceans have shown abrupt
declines in primary productivity, 42\% if accounting only the linear
negative trend; while 252 fish stocks resulted in abrupt declines on
historical data. The literature already showcase past events with
impacts on employment and economic losses on the millions of dollars. So
who is at risk today?

We found that 15\% (16,878) of the vessels tracked in Global Fishing
Watch do actually fish in areas susceptible to marine regime shifts. We
did not find the levels of market concentration found in other economic
sectors such as aquaculture, in fact the ownership structure is quite
sparse with most companies owned by one or two shareholders.
Nonetheless, we find that Norway and China are important actors in
hosting many of the vessels and shareholders in our data, but they are
not strongly involved on foreign investment. Spain, the UK and France
are central countries of origin for foreign investment, while Russia is
a hotspot attracting these investments. The likelihood of these
financial flows seems to be related to differences in government
effectiveness and political stability.

We provided lists of top shareholders and top companies with vessel
ownership fishing in areas prone to marine regime shifts. We discussed
the limitations of our study mainly related to data transparency.
Nonetheless, we hope these lists of companies and shareholders help to
identify actors with agency to manage the risk of marine regime shifts.
We discussed potential interventions that can reduce regime shifts
risks. Redesigning subsidy programs, lobbying company boards to fish in
low-risk areas, or supporting the implementation of international
agreements that take into account potential range shifts and conflicts
are just a few examples of how these actors can incentivise practices
that are aligned with long term sustainability values.

\subsection{Acknowledgements}\label{acknowledgements}

This work has been supported by the project Networks of Financial
Rupture funded by FORMAS grant 2020-00198; and the Finance to Revive
Biodiversity (FinBio) program, financed by Mistra -- the Swedish
Foundation for Strategic Environmental Research (DIA 2020/10), and the
Swedish Research Council (VR grant 2022-04122). Our work benefited from
discussions with Robert Blasiak.

\section*{References}\label{references}
\addcontentsline{toc}{section}{References}

\phantomsection\label{refs}

\onecolumn

\section{Supplementary Material}\label{sec:SM}

\renewcommand\thefigure{S\arabic{figure}}
\renewcommand\thetable{S\arabic{table}}
\setcounter{table}{0}
\setcounter{figure}{0}
\begin{figure*}[ht]
\centering
\includegraphics[width = 7in, height = 3in]{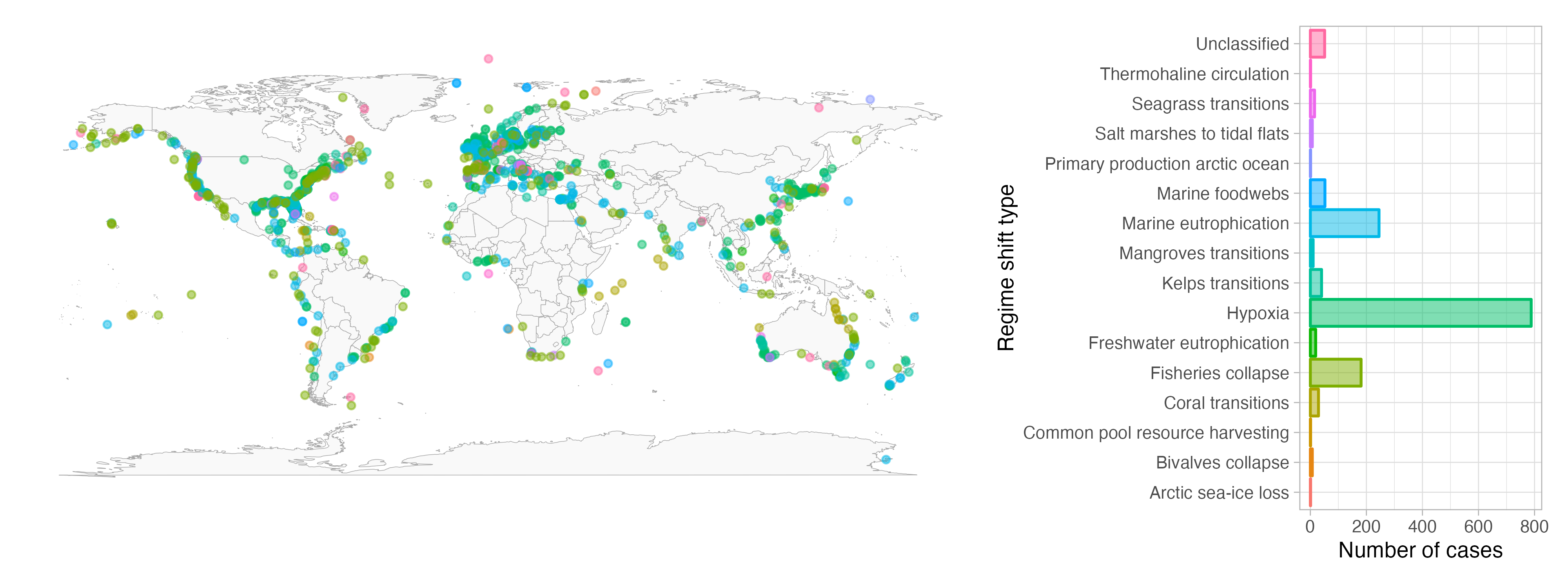}
\caption{\textbf{Empirical evidence of regime shifts}. Over 1440 marine regime shifts are documented on the regime shifts database, 788 cases are of hypoxia, followed by marine eutrophication (245) and fisheries collapse (181). For each dot in the map there is a paper in the scientific literature reporting regime shifts in marine ecosystems. Source: www.regimeshifts.org (beta version at www.regimeshifts.netlify.app).} 
\label{fig:rsdb}
\end{figure*}
\begin{figure*}[ht]
\centering
\includegraphics[width = 7in, height = 7in]{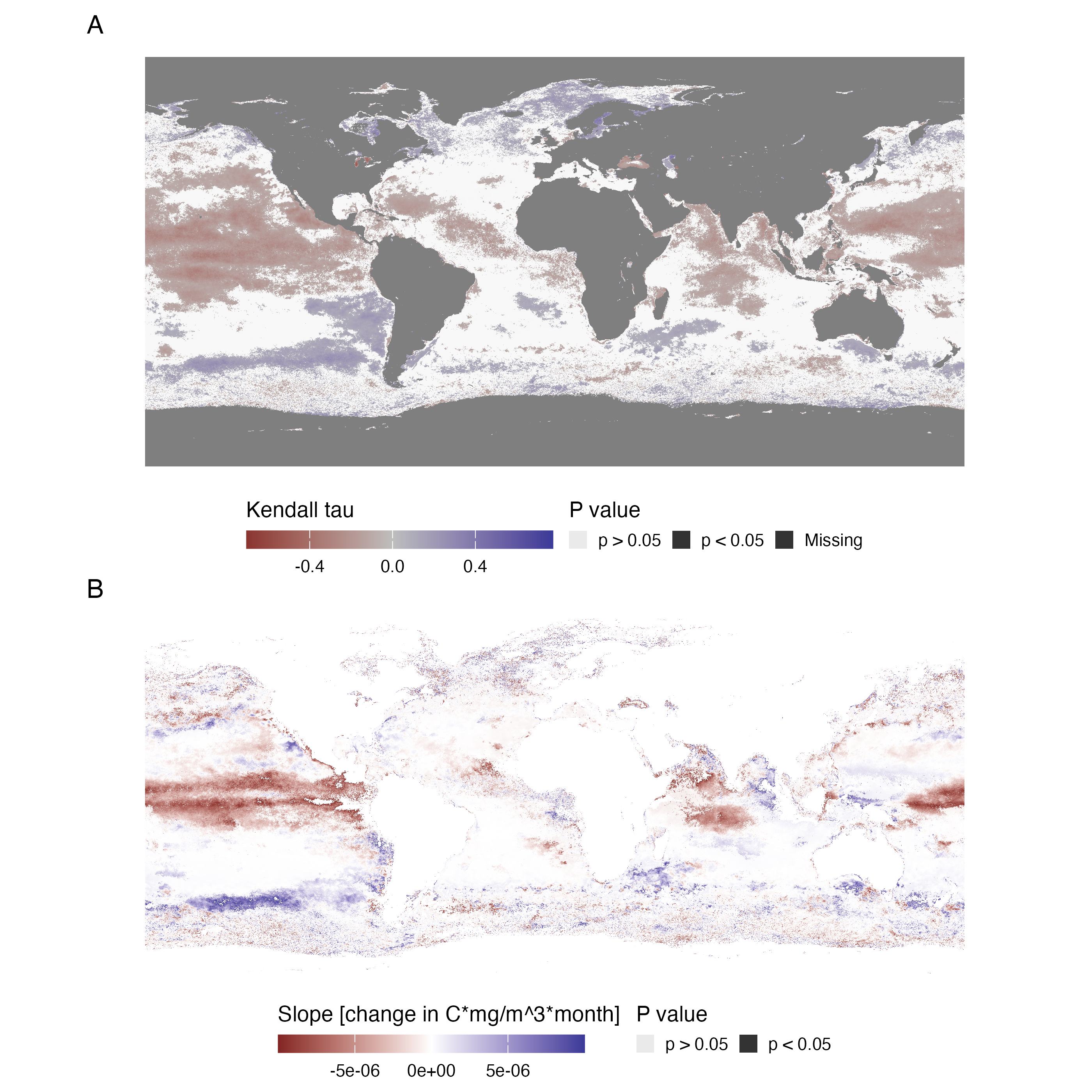}
\caption{\textbf{Linear changes in primary productivity}. We estimated Kendall tau on the full time series of Chlorophyll-A at 9Km resolution. 77.3\% of the ocean shows a slope significantly different from zero, 41.9\% with a declining trend (A). In (B) the same time series have been detrended using seasonal-trend decomposition using LOESS, and a linear model fit on the residuals showing that there are still negative and significant trends in the central Pacific and Indian oceans that are not related to seasonality.} 
\label{fig:slopes}
\end{figure*}
\begin{figure*}[ht]
\centering
\includegraphics[width = 7in, height = 3in]{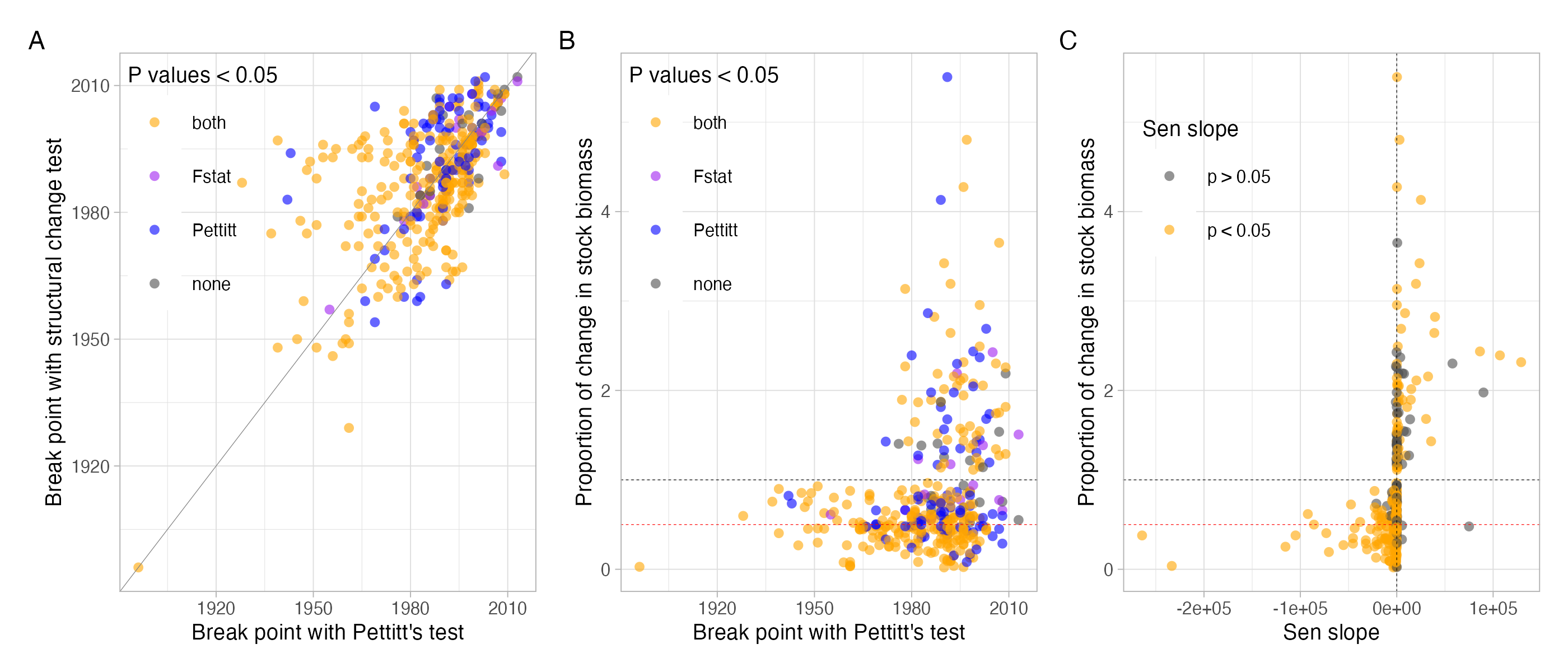}
\caption{\textbf{Linear changes in fish stocks}. We first compared the timing of the break points detected with the Pettitt test and the structural change test (A). Despite both test agree on the existence of the shift, there are some differences in timing. The median decline is 50\% (red dotted line in B-C), and most shifts occurred between the 1970s and early 2000s. A comparison with the Sen slope test show that linear trends agree with the direction of change reported on the abrupt changes, the units of the slope are in biomass.} 
\label{fig:fish_stocks}
\end{figure*}
\begin{figure*}[ht]
\centering
\includegraphics[width = 3.5in, height =2.5in]{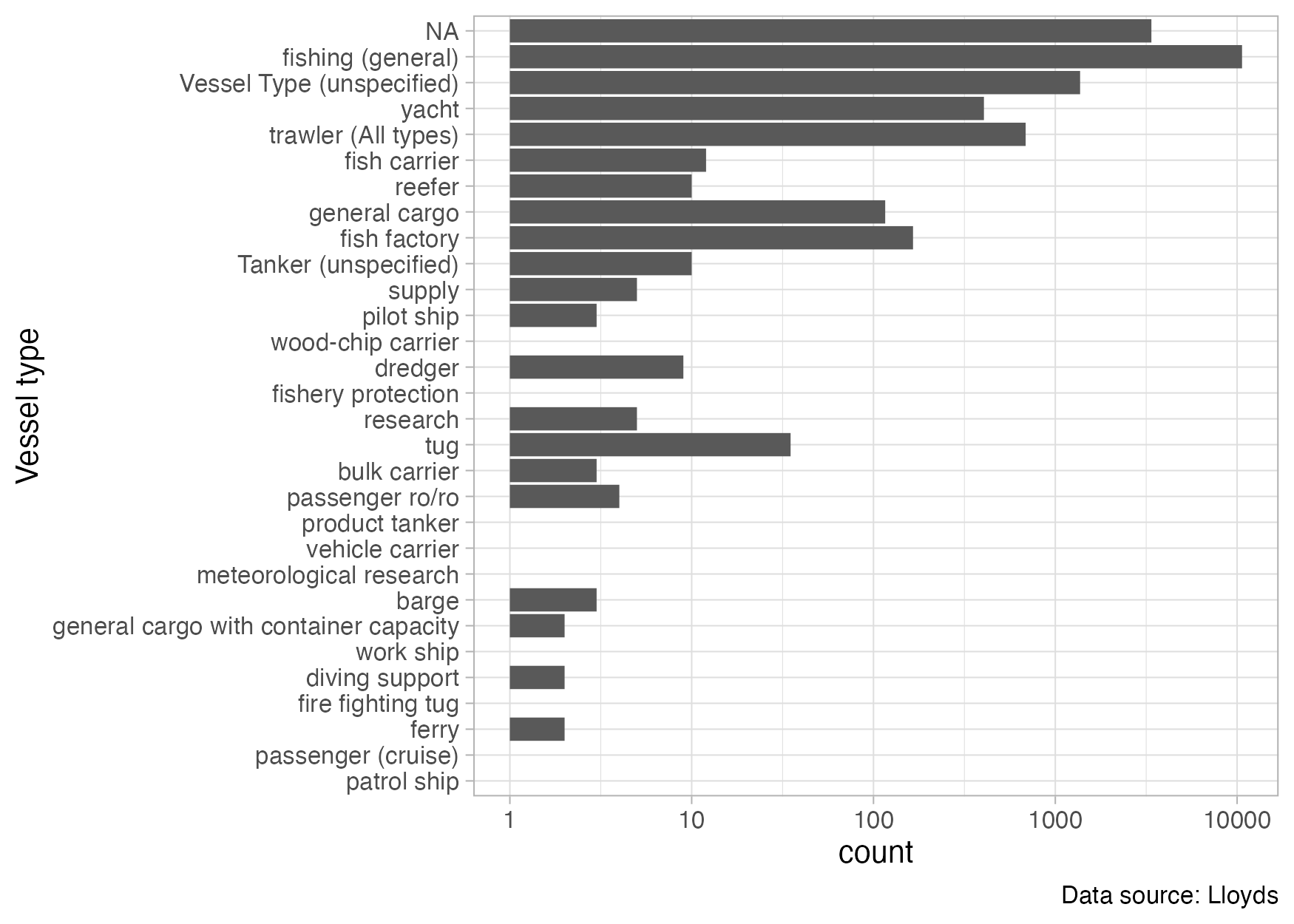}
\caption{\textbf{Vessel types} }
\label{sm:vessels}
\end{figure*}
\begin{figure*}[ht]
\centering
\includegraphics[width = 5in, height =3in]{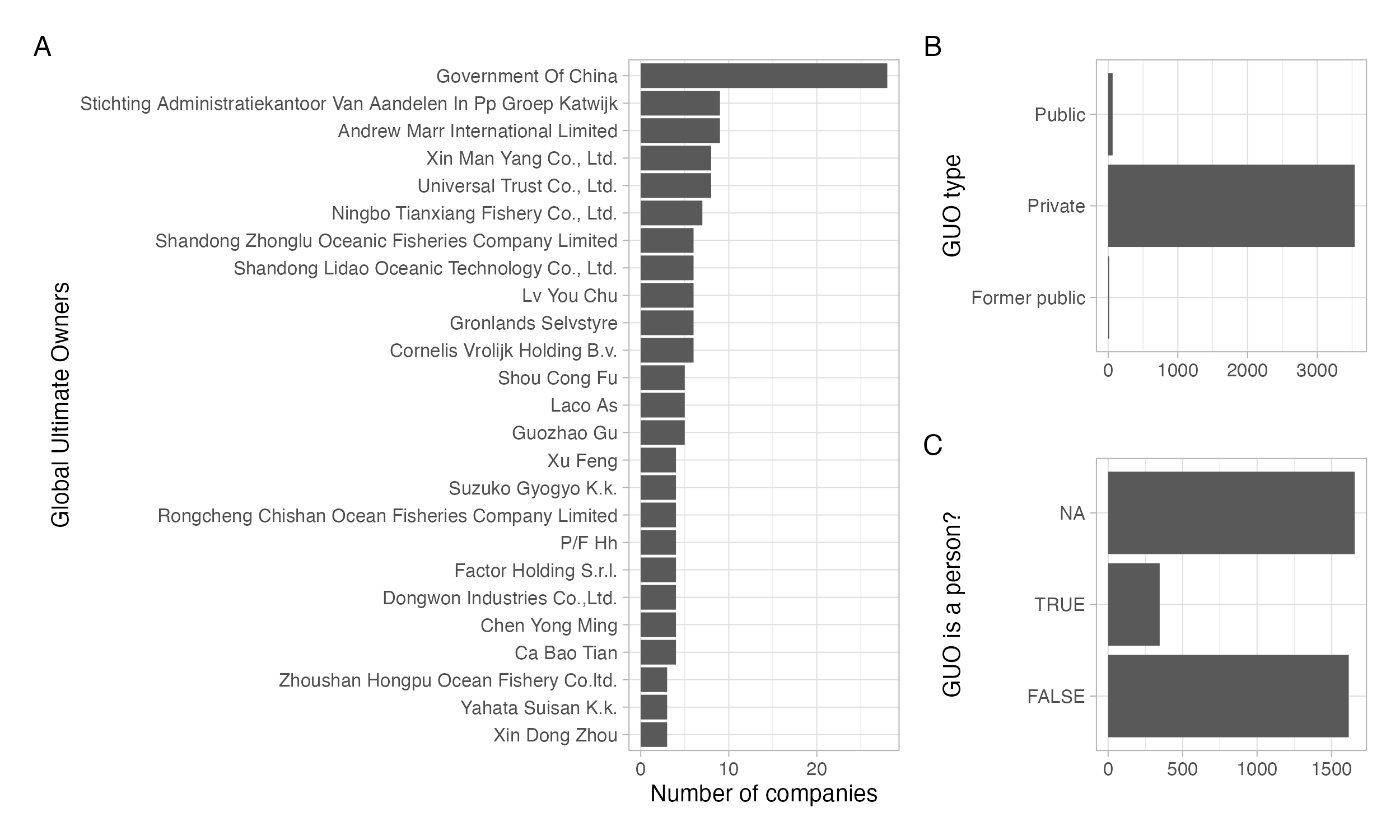}
\caption{\textbf{Global Ultimate Owners} }
\label{sm:guos}
\end{figure*}

\begin{table}[!htbp] \centering 
  \caption{Model statistics} 
  \label{tab1} 
\begin{tabular}{@{\extracolsep{5pt}} cccc} 
\\[-1.8ex]\hline 
\hline \\[-1.8ex] 
 & model & logLik & AIC \\ 
\hline \\[-1.8ex] 
1 & null & -537.22 & 1076.43 \\ 
2 & difference & -527.78 & 1073.56 \\ 
3 & full & -516.75 & 1065.5 \\ 
\hline \\[-1.8ex] 
\end{tabular} 
\end{table}

\end{document}